\documentclass[aps,10pt,superscriptaddress,prl,twocolumn,showpacs]{revtex4}
\usepackage{epsfig,graphics,amssymb,amsmath,subeqnarray}
\begin{document}
\title{Irreversibility and self-organisation\\
 in hydrodynamic echo experiments.}
\author{Gustavo D\"uring}
\affiliation{Laboratoire de Physique Statistique de l'Ecole Normale Sup\'erieure, CNRS UMR 8550-Universit\'e Paris 6-Universit\'e Paris 7; 24, rue Lhomond, 75005 Paris, France}
\author{Denis Bartolo}
\author{Jorge Kurchan}
\affiliation{PMMH-CNRS UMR 7636-ESPCI-Universit\'e Paris 6-Universit\'e Paris 7; rue Vauquelin 10, 75231 Paris, France}
\begin{abstract}
We discuss the reversible-irreversible transition in low-Reynolds
hydrodynamic systems driven by external cycling actuation.  We
introduce a set of models with no auto-organisation, and show that a
sharp crossover is obtained between a Lyapunov regime in which any
noise source, such as thermal noise, is amplified exponentially, and a
diffusive regime where this no longer holds. In the latter regime, groups of
particles are seen to move cooperatively,  yet no spatial organization occurs.
\end{abstract}
\maketitle
 The behavior of a system that retraces its steps after a reversal
in its dynamics -- the echo protocol -- 
 has been of long-standing  interest.
From a fundamental perspective, the question of how
   macroscopic irreversibility arises from the reversible dynamics
of the microscopic components led to one of the central debates
on the foundations of statistical mechanics between Loschmidt and 
Boltzmann~\cite{swendsen}.  Loschmidt argued that reversing the
velocities of all the particles in a box should allow for the system
to return to its initial position, thus invalidating the notion of 
an arrow of time. 
In later years,  echo experiments became  practical tools as  Nuclear 
Magnetic Resonance and Neutron
Spin-echo  techniques to probe irreversible
microscopic events in soft materials, and has also, 
 as the name would suggest, numerous applications in acoustics \cite{fink}. From a numerical perspective, Levesque and Verlet showed how rounding off errors suffice to destroy reversibility in classical Hamiltonian systems~\cite{verlet}.
A conceptually closely related situation, in the field of hydrodynamics,
is the classic Taylor experiment
to denmonstrate
 the prefect reversibility of viscous fluid flows~\cite{taylor}. It
consists of adding a drop of dye to a viscous fluid in the gap between two
concentric cylinders.  The drop is then strongly stretched by
 turning the inner cylinder. By subsequently imposing the
reverse rotation, mixing is not enhanced: on the contrary, the initial
spherical shape of the drop is  recovered. 
A rather spectacular elaboration of this experiment 
 has been recently performed by Pine et al.~\cite{pine}, who substituted
the ink droplet with a high
volume fraction of non brownian beads. Although  shear
flow induces an effective diffusion of the particles in
the suspension due to hydrodynamic interactions~\cite{eckstein77,acrivos87},
 the complicated
particle trajectories thus generated during half a cycle
should in principle be retraced in the second half.
 By measuring the net particle
displacement after each back and forth  cycle, Pine et al.  uncovered
a remarkable transition from a reversible situation
in which particles do retrace their steps to a regime 
 obtained  above a critical strain
amplitude, in which reversibility is lost.

Two different but related 
questions immediately arise: {\em (i)} what
is the origin of irreversibility, and {\em (ii)} is the transition a sharp one,
and if so, what is its nature?
In a recent paper, Corte et al proposed that the key point could be
the close particle encounters, wehre non-hydrodynamic -- and hence
non-reversible --
 contact interactions (Van der Waals,  mechanical friction)  act.  In one
stroke this provides both
 an explanation for irreversibility and a mechanism for a sharp
transition: because particles move away from situations that generate 
diffusion, they self-organise in subsequent cycles into a  configuration
in which they avoid encounters, and once this is done irreversible diffusion 
stops. Above a certain percolation-like transition the system is no
longer able to find such a configuration without collisions,
 and irreversibility persists in time.
This scenario was further studied analytically by Menon and Ramsawami 
\cite{sriram}, who made the
relation to percolation  quantitative.

In this paper we test the opposite scenario: we study models that,
 by construction, cannot lead to self-organization. We assume
that small irreversible perturbations, such as Brownian motion, exist
all along and are amplified by chaoticity.  This will take us to a
situation that is very close to that of echoes in classical
Hamiltonian systems.  We shall show that a sharp increase of particle
diffusion can be observed in such an "hydrodynamic echo experiment"
even if no real dynamic phase transition and no self-organization
process occurs.  Although it is likely that the contact-induced
self-organization scenario is relevant for the experiments in
Ref.~\cite{pine}, we believe that other experimental situations (and
perhaps even the numerical hydrodynamic simulation in~\cite{pine}) may
well correspond to simple chaotic amplification.

Consider $N$ force free particles immersed in a viscous incompressible
fluid bounded between  two concentric cylinders, the inner one 
being able to rotate at frequency $\omega(t)$. 
 For spherical particles and creeping flows,
the particle velocities  are linearly related to $\omega$~\cite{pozrikidis}:
\begin{equation}
\dot{\bf{x}}_a = M_a({\bf{x}}_1,...,{\bf{x}}_{N}) \omega,
\end{equation}
where the ${\bf{x}}_a$ and $\dot{\bf{x}}_a$ are the particle center
positions and velocities. We shall henceforth set $|\omega|$ to one.  The
mobility coefficients $M_a$ are complex  vectorial functions,
obtained in principle by solving the stationary Stokes problem with
the instantaneous boundary conditions, and eliminating the angular
velocities using the fact that the particles are torque-free. The
evolution is reversible, if we make half a period $T/2$ with ${ \omega}$
and subsequently $T/2$ with $-{ \omega}$,
 each particle retraces its steps
over a cycle as in Taylor's experiment. Next, consider the effect of a
noise $\eta_a(t)$ of small amplitude, $\epsilon$, acting on the
particle $a$ that we shall assume white but in general with spatial
correlation $\langle \eta_a(t) \eta_b(t')\rangle = 2 \epsilon^2
\delta(t-t'){\tilde M}_{ab}({\bf{x}_a, \bf{x}_b})$.  In the
specific case of the thermal noise, ${\tilde M}_{ab}({\bf{x}_a,
\bf{x}_b})$  is equal to the mobility matrix, such as to respect
the Fluctuation-Dissipation theorem~\cite{Pusey}.  The equation of
motion for the $N$ particles is
\begin{equation}
\dot{\bf{x}}_a = \left\{
\begin{array}{ccc}
+M_a({\bf x}_1,...,{\bf x}_{N})+\eta_a(t), &{ \rm if} & 0<t<T/2 \\
-M_a({\bf x}_1,...,{\bf x}_{N})+\eta_a(t), & { \rm if} & T/2<t<T
\end{array}
\right.
\label{evo}
\end{equation}

For $\epsilon \neq 0$, given that the noise is different in the two
semicycles, the trajectory does not retrace its steps exactly.  How
much does a trajectory deviate from the initial position after a
cycle?  Importantly, the invariance upon time reversal of
Eq.~\ref{evo} tells us that this question is equivalent to asking how
much two trajectories, starting from the same position at mid-cycle
diverge under the effect of different noise realisations. 
Let us first
make this discussion for small noise, and for large cycle times. In
this limit, we can express the separation, $\Delta$, of the particles
between the two trajectories in the $3N$ dimensional phase space in
terms of the Lyapunov exponents~\cite{prozen}
\begin{equation}
 \Delta \propto  e^{\lambda^{\rm traj}\;  T/2 }
\label{lyap}
\end{equation}
where $\lambda^{\rm traj}$ is the largest Lyapunov exponent.
  An important remark that should be
made is that, because  $\lambda^{\rm traj}$ is a function of the  trajectory,
depending on the specifics of the driving, it may or may not coincide
with the one of a randomly chosen trajectory after a few cycles. For
instance, the system may drift away from the regions with high
Lyapunov exponents and converge to a subset of the phase space
corresponding to smaller diffusivity. We will refer to such a
phenomenon as {\em self-organization}. Clearly, this definition of
self-organization also holds for larger noise levels, when the
Lyapunov linearization no longer holds.

We now restrain ourselves to a limiting case in which
self-organization does not happen.  Let us consider the limit of
pointwise particles, or equivalently of very dilute suspensions. In
this limit, the particles behave like fluid tracers. It thus follows
from the (fluid) incompressibility conditions that the mobility
coefficients obey:
\begin{equation}
\sum_a {\nabla}_a {{M}}_a =0,
\label{virtue}
\end{equation}
with $\nabla_a\cdot\equiv\partial\cdot/\partial_{\bf{x}_a}$.  This
implies that the probability distribution $P({\bf x_1},...,{\bf x_N})$
converges to the flat measure. Indeed, during each half-cycle, $P$
evolves according to the Fokker-Plank equation \cite{Pusey},
\begin{equation}
\dot P = \sum_{ab} \nabla_b \left\{ \epsilon^2   
{\tilde M}_{ba} \nabla_a \pm \delta_{ab} M_a \right\}P,
\end{equation}
which admits a constant function as a solution by virtue of
Eq.~\ref{virtue}.  In this case, there can be no self-organisation. In
fact, if the initial positions where chosen with flat probability, the
probability distribution does not evolve and no {\em static} one-time
correlation function depends on the cycle number.  {\em Note that
there are two independent issues concerning external forces and
interparticle potential interactions: (i) whether they break time
reversal, and (ii) whether  they  preserve or not the
flat measure.} Needless to say, in a system with divergenceless forces
there can be no structural phase transitions, other than those of
equilibrium hard spheres.  Furthermore, the separation of nearby
trajectories corresponds to that of an equilibrium unbiased system,
and in particular Lyapunov exponents are the 'typical' ones sampled in
equilibrium.  Because of the obvious analogy with the well-studied
classical and quantum mechanical problem, we shall call this the
hydrodynamic Loschmidt-echo situation.

To stress the analogy, we shall in this paper substitute hydrodynamic
evolution, Eq.~\ref{evo}, by simple Hamiltonian dynamics, in the
presence of random noise, such that it conserves the microcanonical
distribution.  Our first example is very similar to the original
Loschmidt experiment: we consider  particles in two dimensions with power
law $r^{-3}$ interaction, perturbed by small energy-conserving noise (the choice of a long range interaction, by analogy with hydrodynamic coupling, is intended to limit the role of particle encounters). 
  The second example,
 which allows us to go to large sizes and times, is a system of
 coupled simplectic maps \cite{kaneko,kantz}.
 
\begin{figure}
  \begin{center}
    \includegraphics[width=\columnwidth]{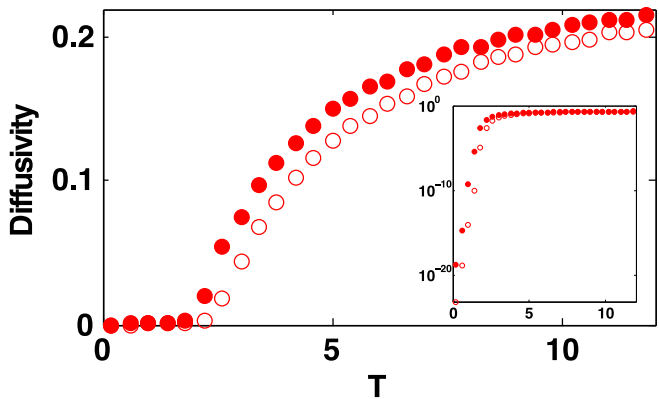}
    \caption{Diffusivity $D$ versus period $T$ for the $2$D system composed of $N=250$ particles. Over each half-cycle, the particles positions evolve according to: $\ddot{\bf x}_a=-\sum_b \alpha|{\bf x}_b-{\bf x}_a|^{-4}+\epsilon_a$, where the $\epsilon_a$ are uncorrelated Gaussian white noises of variance $\epsilon$ and $\alpha=2\,10^{-6}$. The box size is set to $1$  and  the energy is chosen equal to $E\approx87$. The diffusivity is averaged over $100$ cycles.  The open symbols correspond to the zero additional noise limit, $\epsilon=0$, yet the trajectories are not reversible. This is due to the (irreversible) numerical rounding off~\cite{verlet}. The filled symbols correspond to $\epsilon=(1/\sqrt{2})10^ {-7}$. Inset: same plot  in log scale.}
    \label{fig1}
  \end{center}
\end{figure}

Consider first the two-dimensional system. We perform direct followed
by time-reversed evolution, after a velocity-reversal in mid-cycle. If
the system is started in a thermalized microcanonical configuration,
then configurations are statistically distributed in the same way at all times.
 Figure~\ref{fig1} shows the average quadratic
dispacement in one cycle in terms of the cycle time, for two values of
noise amplitude.
The shape of the curve is very similar to the one reported for sheared
suspensions and for superconducting vortices~\cite{pine,reichhardt}.
 The mean squared displacement first
increases exponentially with $T$ and then saturates to a constant
value. 

This behavior can be explained as follows. At short periods,
back and forth trajectories differ only slightly, and the Lyapunov
linearization applies. Because, in the thermodynamic limit, there is a
stable Lyapunov density function and  two subsequent Lyapunov
exponents differ by $O(1/N)$ 
there are then $O(N)$ largest Lyapunov
directions
which contribute to the instability, since  the separation
they induce are indeed comparable
at any finite time, see~\cite{prozen,Ruffo}. This implies that the
separation per particle is stable in the thermodynamic limit,
everything else  (period, particle and energy
density) being kept equal.
At times $T^*$ such that $\epsilon e^{\lambda T^*}$ is a sizable
fraction of the interparticle distance, the linearization breaks
down~\cite{Vulpiani}.  Moreover, for $T>>T^*$  the system
completely loses memory in a cycle, and the separation becomes
diffusive rather than exponential.  Note, first, that the apparent
transition time $T^*\sim \lambda^{-1} \ln \epsilon$ depends very
weakly on the irreversible noise amplitude, and, second, that it has a well-defined (large $N$) thermodynamic limit.  In turn, the main
features of the fluctuations of the stroboscopic positions with $T$
are insensitive to the specific process that breaks the time reversal
symmetry.   Note also that the saturation of the actual particle {\em self}-diffusion occurs 
when the typical displacement of order of the box size.  If we now make
plots with the y-axis in linear scale and normalised by the
saturation, the curve will have a sharp inflection, Figure 1.
In this sense, and probably only in this sense, there is a sharp
transition without self-organisation.

\begin{figure}
  \begin{center}
    \includegraphics[width=\columnwidth]{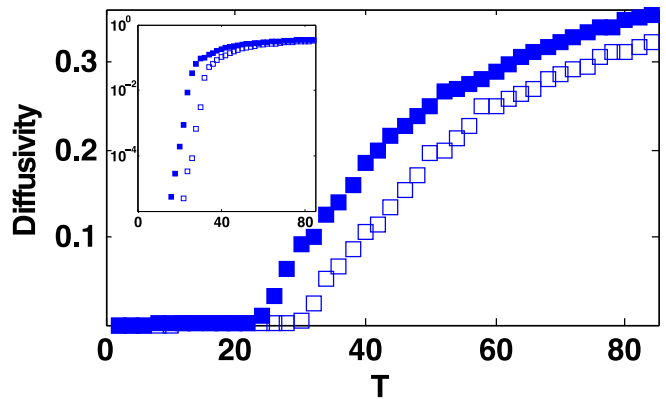}
    \caption{
Diffusivity, $D$, versus period, $T$, for $N=7000$ coupled maps.  The diffusivity has been  averaged over $500$ cycles. Open (resp. filled) symbols correspond to a noise amplitude $\epsilon=(1/\sqrt{2})10^ {-12}$ (resp. $\epsilon=(1\sqrt{2})10^ {-10}$). Inset: same plot  in log scale.}
    \label{fig2}
  \end{center}
\end{figure}

\begin{figure}[t!]
  \begin{center}
     \includegraphics[width=0.87\columnwidth]{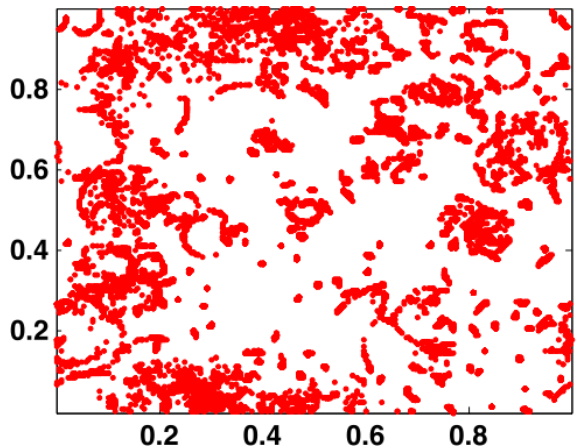}
    \caption{Stroboscopic  plot of the particles position after  each cycle for the 2D system of interacting particles. Plot after $100$ cycles for a period  $T=2$, same parameters as in Figure~1 (open symbols).}
    \label{fig2.1}
  \end{center}
\end{figure}

\begin{figure}[t!]
  \begin{center}
    \includegraphics[width=0.87\columnwidth]{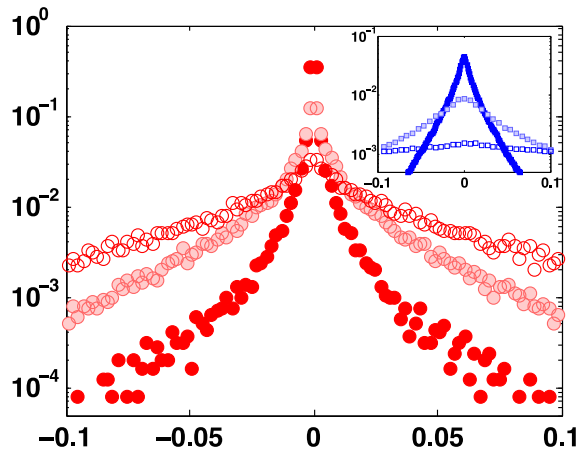}
    \caption{
 Histograms of particles diffusivity for $T=2$ (dark red circles), $T=2.2$ (light red circles) and $T=2.4$ (open circles). Same parameters as in Figure~1 (open symbols).{\bf Inset}: Same histograms for the coupled maps with $\epsilon=(1/\sqrt{2})10^{-12}$ and $T=28$ (dark blue squares), $T=30$ (light blue squares) and $T=32$ (open squares).}
    \label{fig3}
  \end{center}
\end{figure}

In figure~\ref{fig2} we show the same results for a toy model: a time-discrete Hamiltonian system of $N$ coupled maps constructed by the composition of the two following one step iterations and their time-reversed~\cite{kantz}:
\begin{equation}
\left.
\begin{array}{l} p'_{i}=p_i+\epsilon\, {\xi}_{p_i}\\
q'_{i}=q_i+p_i+\epsilon\,{\xi}_{q_i}
\end{array}
\right\}  {\rm mod}\, 1
\end{equation}
\begin{equation}
\left.
\begin{array}{l} 
p'_i=p'_i+\frac{1}{\sqrt{N-1}}\sum_{j=1}^{N}\sin{\left[2\pi(q'_i-q'_j)\right]}\\
q''_i=q'_i
\end{array}\right\}  {\rm mod}\,1
\end{equation}
where $q'_i(t)\equiv q_i(t+1)$, $p'_i(t)\equiv p_i(t+1)$. ${\xi}_{p_i}$ and ${\xi}_{q_i}$ are Gaussian white noise of variance one. Again the fluctuations of the stroboscopic
position in the $(q,p)$ plane display the same features. 
Note that for this simple model
we achieved $N=7000$ and have checked that the curves for different (large) $N$ coincide within numerical error and did not observe any sharpening of
the  crossover. 
Figure \ref{fig2.1} shows the motion of particles just above the
mobility time. Surprisingly enough, even in the absence of
self-organization the particle system  seems 
not entirely devoid of spatial features.  At a given time, there are
three types of particles: almost immobile, those that move {\em back
and forth} along essentially one-dimensional trajectories, and those
that have a more typical diffusive character. This latter group --
which well above the crossover becomes dominant -- looks spatially
correlated, suggesting that these are the particles that interact the
most along trajectories.
In Fig. \ref{fig3} we show histograms of particle diffusivity in one cycle.
The distribution seems exponential, and there are relatively many 
highly diffusive particles at any time. The observation of some degree of dynamic heterogeneity does not contradict the fact that static correlations are those of the equilibrium system, since this  involves a purely dynamic correlation, the spatial correlations between particle {\em displacements} in a cycle.

The scenario discussed here is by construction the opposite of the one
considered in Refs \cite{chaikin,sriram}, which involves a dramatic
change in the stationary end-of-cycle distribution near the
transition, completely absent here.  In Ref.~\cite{chaikin}, the time
for reaching the stationary value of diffusivity was plotted, showing
a peak near the transition strain, an evidence of organization.  The
situation could however be mixed: consider a system with perfect
time-reversal broken by a constant Brownian noise, so that the
flat distribution is preserved.  Suppose we switch on forces that
violate the {uniform} distribution: this indeed is what happens in the
hydrodynamic case as  the volume fraction is increased.  The
system will now self-organize into a distribution that, unlike the
case we studied here, will depend on strain, noise level and particles
volume fraction.  If the effect is weak, we do not expect that a sharp
transition will immediately arise, so that from this point of view the
scenario is still the one we discussed here, in spite of having a
certain degree of auto-organisation.  
In order to further emphasise
the experimental relevance of the hydrodynamic echo scenario, we
compute the crossover shear amplitude $\gamma^*$
which would correspond to the experimental conditions reported
in~\cite{pine,chaikin}. The noise source is supposed to be the weak
Brownian motion of the suspended particles. $\epsilon$ is thus
approximated by the distance over which a single particle
diffuses over half a cycle: $\epsilon\equiv\sqrt{DT}$, $D=k_{\rm
B}T/(6\pi\eta a)$ with $\eta=3$Pa$.$s the fluid viscosity and
$a=115\mu$m the particles mean radius~\cite{chaikin} (Note that we have 
neglected hydrodynamic correlations of the noise). 
The interparticle distance is of order of $\Delta\sim
a/\phi^{1/3}$. To compute $\gamma^*$, we also need the Lyapunov
exponent of the concentrated particle suspension. This value has been
computed numerically  in~\cite{acrivoschaos}, for $\phi=0.4$:
$\lambda \sim 0.6\dot\gamma$ where $\dot \gamma\equiv2\pi\gamma/T$ for sinusoidal cyclic shear. We thus infer a critical
shear amplitude of: $\gamma^*\sim3$ which is  fairly close to the
experimental value: $\gamma^*_{\rm exp} \sim 1$ given the crudeness of
 of our numerical estimates. 
This predictions suggest that there are very probably many
situations intermediate between the pure Loschmidt echo crossover
discussed in this paper and percolation-like transition.  In such
cases, the relative importance of the two effects cannot be solely
revealed by the measurement of the effective diffusivity of the
particles and may be harder to assess experimentally.

J.-L. Barrat, P. Chaikin, L. Cipelletti, L. Corte, G. Menon, D. Pine, S. Ramaswamy and especially D. Levine are acknowledged for valuable comments and discussions. G. D. was supported by a CONICYT grant.


\begin{thebibliography}{99}

\bibitem{swendsen} R. H. Swendsen {\it Am. J. Phys.} {\bf 76}, 643 (2008) 

\bibitem{fink} M Fink, {et. al.}, {\it Rep. Prog. Phys.} {\bf 63}, 1933 (2000)

\bibitem{verlet}
D. Lesveque and L. Verlet, {\it J. Stat. Phys.} {\bf 72}, 519 (1994)

\bibitem{taylor}
G. I. Taylor, National Committee for Fluid Mechanics 
Films, Education Development Center, Newton, MA (1966)

\bibitem{pine} 
D. J. Pine, J. P. Gollub, J. F. Brady, A. M. Leshansky, {\it Nature} {\bf 438}, 997 (2005)

\bibitem{eckstein77} 
E. C. Eckstein, D. G. Bayley and A. H. Shapiro, {\it J. Fluid. Mech.} {\bf 79}, 191 (1977)

\bibitem{acrivos87}
D. Leighton and D. Acrivos  {\it J. Fluid. Mech.} {\bf 177}, 108 (1987).

\bibitem{chaikin} Laurent Corte, P. M. Chaikin, J. P. Gollub and
  D. J. Pine, {\it Nature Physics} {\bf 4}, 420 (2008)

\bibitem{sriram} G. I. Menon and S. Ramaswamy, {\it cond-mat} arXiv:0801.3881v1 (2008)

\bibitem{pozrikidis} 
J. Happel and H. Brenner, {\it Low Reynolds Number Hydrodynamics}, Springer (1981) 

\bibitem{Pusey} P Pusey, in {\em Liquids, Freezing and the glass transition}
Edited by JP Hansen, D Levesque and J Zinn-Justin, Elsevier Amsterdam (1991)

\bibitem{prozen} G. Veble and T. Prosen, {\it Phys. Rev. E}, {\bf 72}, 025202(R) (2005)

\bibitem{kaneko} T Konishi and K Kaneko,  {\it J. Phys. A: Math. Gen.} {\bf 25}, 6283 (1992)

\bibitem{kantz} E. G. Altmann and H. Kantz, {\it Europhys. Lett.} {\bf
  78}, 10008 (2007)
\bibitem{reichhardt} 	
N. Mangan, C. Reichhardt, and C. J. Olson Reichhardt, 
{\it Phys. Rev. Lett.} {\bf 100}, 187002 (2008)
\bibitem{Ruffo} 
R Livi, A Politi and S Ruffo, {\it J. Phys. A: Math. Gen.} {\bf 19}, 2033 (1986) 
\bibitem{Vulpiani} 
G. Boffetta, M Cencini, M Falcioni and A Vulpiani
{\it Phys. Rep.}  {\bf 356}, 367 (2002) 
\bibitem{acrivoschaos}
G. Drazer, J. Koplik, B. Khusid, A. Acrivos, {\it J. Fluid. Mech.} {\bf 460}, 307 (2002)
\end{thebibliography}
\end{document}